\documentclass{PoS}

\usepackage{subfigure}

\def\lumi{33~\rm{pb}$^{-1}$}

\def\mdv{\ensuremath{m_{\rm DV}}}
\def\ntdv{\ensuremath{N^{\rm trk}_{\rm DV}}}

\def\beq{\begin{equation}}
\def\eeq{\end{equation}}
\def\beqa{\begin{eqnarray}}
\def\eeqa{\end{eqnarray}}

\title{Search for R-parity violating supersymmetry with the ATLAS detector}

\ShortTitle{Search for RPV SUSY with ATLAS}

\author{\speaker{Paul D. Jackson, on behalf of the ATLAS Collaboration}\\
        SLAC National Accelerator Laboratory\footnote{now at the University of Adelaide}\\
        E-mail: \email{paul.jackson@cern.ch}}

\abstract{R-parity violation in supersymmetry gives rise to many unique experimental signatures. We describe searches with the 
ATLAS detector for supersymmetry with R-parity violating decays. Examples include searches for resonant sneutrino decays 
to an electron and a muon, and displaced vertices arising from the late decays of heavy objects with a muon in the final state. 
The most recent results on these channels are presented based on data recorded with the ATLAS detector in 7~TeV $pp$ collisions at
the CERN Large Hadron Collider in 2010 and 2011.}

\FullConference{The 2011 Europhysics Conference on High Energy Physics-HEP 2011,\\
		July 21-27, 2011\\
		Grenoble, Rh\^{o}ne-Alpes France}

\begin{document}

\section{Introduction}

An essential aspect of all supersymmetric (SUSY) models is the R-parity quantum number 
$R_p = (-1)^{3B+L+2S}$, where $B$ denotes the baryon number, $L$ the lepton number
and $S$ the spin of a particle. In the most general formulations of supersymmetric
theory $R$-parity is not conserved, allowing couplings between two ordinary fermions
and a squark ($\tilde{q}$) or slepton ($\tilde{\ell}$). 
The conservation of R-parity forbids strong baryon and lepton number violation simultaneously
and leaves the lightest particle as a natural dark matter candidate.
R-parity violating (RPV) couplings are allowed in scenarios which forbid either the baryon or lepton 
number violating terms. Here, the dark matter candidate is unstable due to new channels being opened.

This note reports on two searches performed by the ATLAS collaboration~\cite{bib:atlas} which can be interpreted in
terms of constraints on RPV SUSY~\cite{bib:rpv}. We search for a resonant structure decaying, into oppositely charged
electron and muon pairs. A separate analysis searches for displaced vertices, potentially arising from the decay of heavy particles
with lifetimes between picoseconds and nanoseconds.

\begin{figure}[htbp]
  \subfigure[$\tilde{\nu}_{\tau}\rightarrow e\mu$ decay]{\label{fig:sneutrino}
    \includegraphics[width=0.42\textwidth]{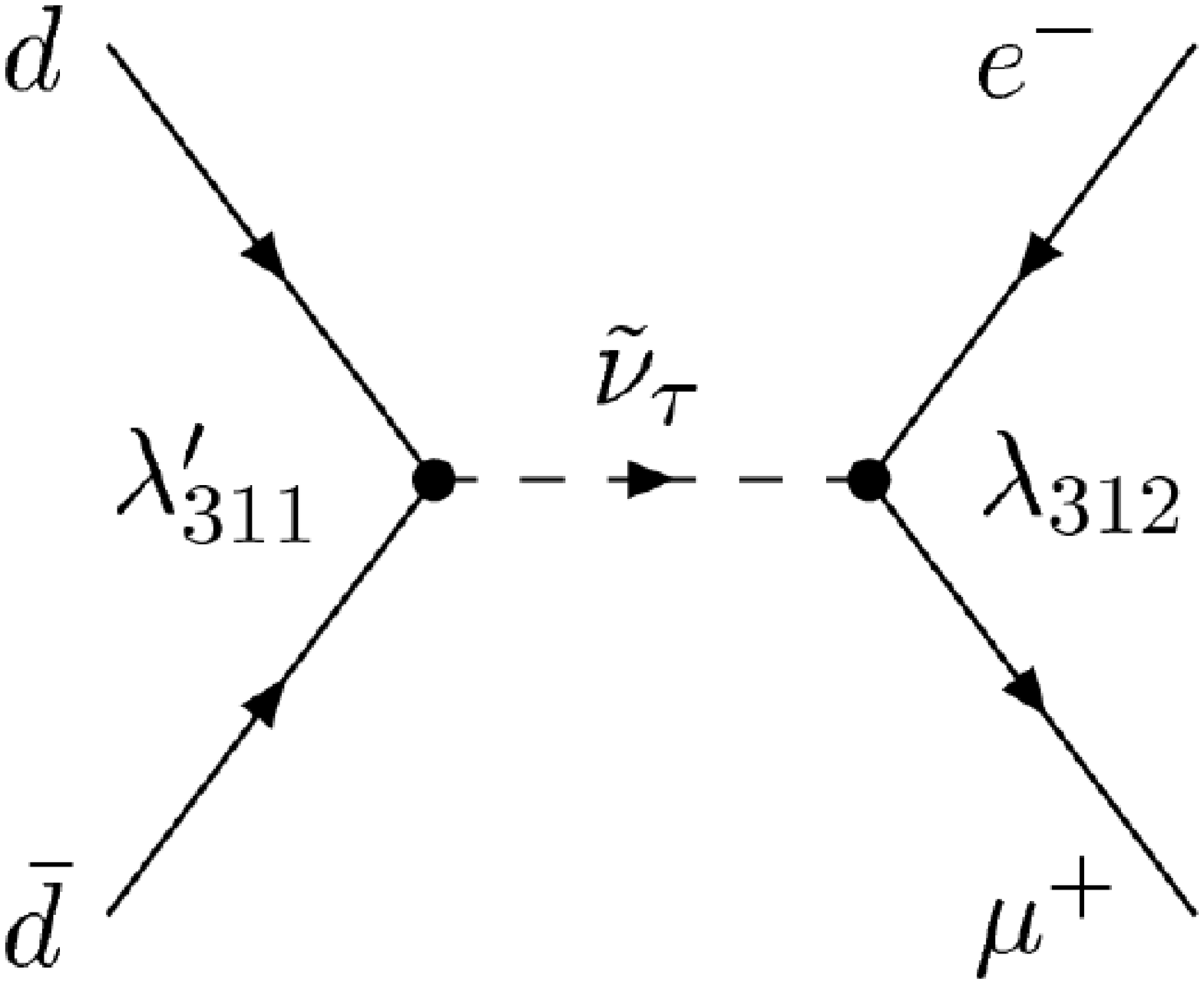}}
  \subfigure[$\tilde{\chi}^{0}$ decaying into a muon and two jets, via a virtual $\tilde{\mu}$, with RPV coupling $\lambda^{'}_{2ij}$.]{\label{fig:dispVert}
    \includegraphics[width=0.42\textwidth]{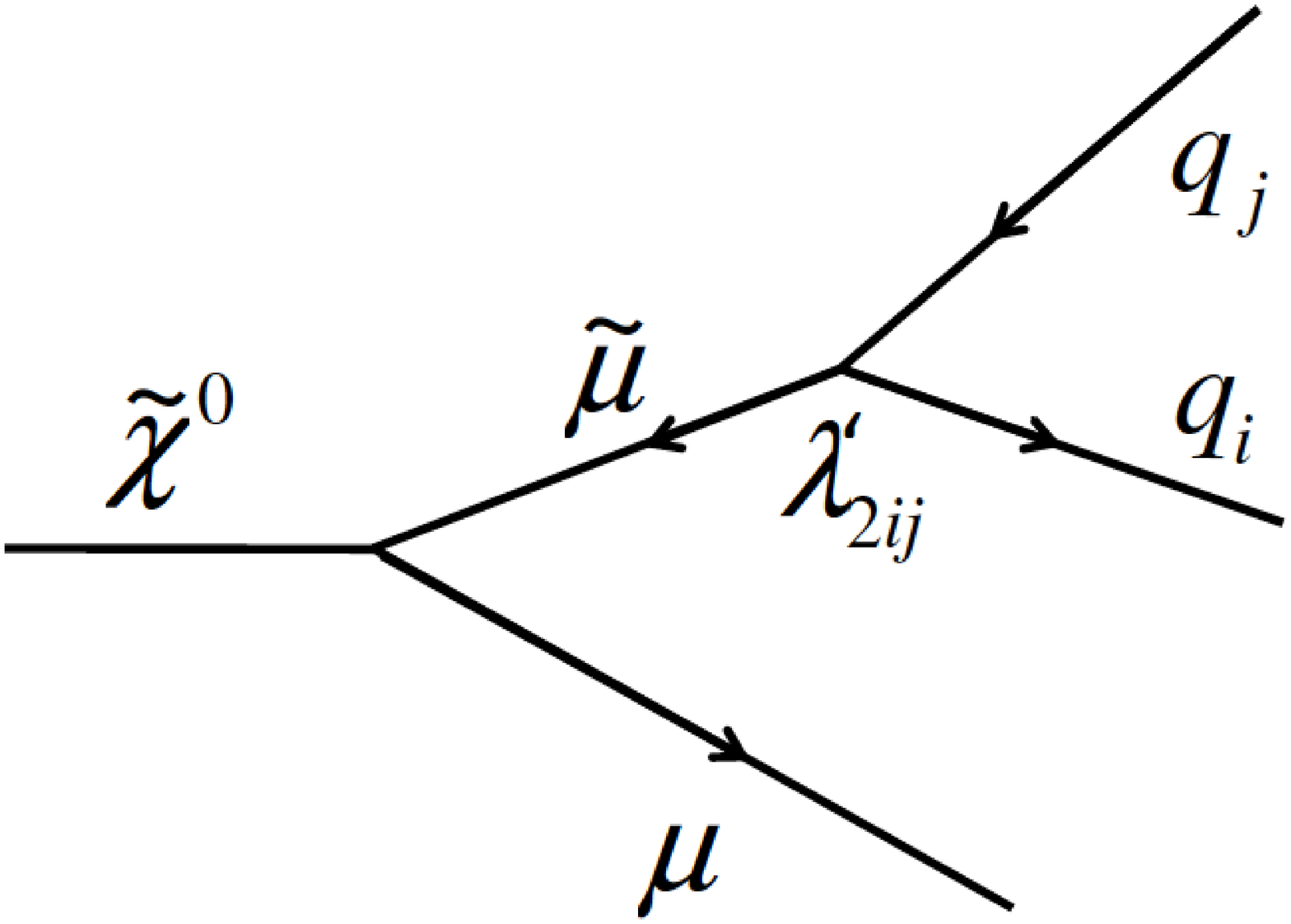}}
 \caption{Feyman diagrams for the two RPV processes under consideration.}
\label{fig:FeynmanDiagrams}
\end{figure}

\section{e-$\mu$ resonance}

By fixing all RPV couplings except $\lambda^{'}_{311}$ ($\tilde{\nu}_{\tau}\rightarrow d\bar{d}$)
and $\lambda_{312}$ ($\tilde{\nu}_{\tau}\rightarrow e\mu$) to zero, and assuming that 
$\tilde{\nu}_{\tau}$ is the lightest supersymmetric particle, the contributions to the 
$e\mu$ final state originate from the $\tilde{\nu}_{\tau}$ only~\cite{bib:emuPaper}.
To select $e\mu$ candidates, the electron is required to have transverse momentum ($p_{\mathrm{T}})~>$~25~GeV, to have  pseudorapidity
$|\eta|~<~1.37$ or $1.52~<~|\eta|~<~2.47$ and satisfy shower shape, track quality and matching criteria.
The muon candidate must be reconstructed in both the inner detector and muon spectrometer, have
$p_{\mathrm{T}}~>$~25~GeV and $|\eta|~<~2.4$, the inner detector track must be isolated from other tracks.
The electron and muon must be separated by $\sqrt{(\Delta\eta^2 + \Delta\phi^2)}~>~0.2$.
The $e\mu$ candidates must be composed of exactly one electron and one muon of opposite charges. 

Standard Model backgrounds to the search can arise from two categories: those with a real final state
$e\mu$ pair ($Z/\gamma^{*}\rightarrow\tau\tau$, $t\bar{t}$, single top, $WW$, $ZZ$ and $WZ$), and processes
which give fake backgrounds ($W/Z+\gamma$, $W/Z+$jets and multijet events with photons or jets
reconstructed as leptons). All processes listed in the first category, along with photon related backgrounds,
are estimated from MC simulation (described elsewhere~\cite{bib:atlasSimulation}).
The remaining fake backgounds are described using a 4$\times$4 matrix method described in~\cite{bib:emuPaper}. 
The lepton definitions are loosened to allow events to be classified based on whether they pass or fail the 
loose and standard requirements. These can then be used to quantify the overall lepton efficiencies 
and jet fake backgrounds.

\begin{figure}[htbp]
  \subfigure[Observed and predicted $e\mu$ invariant mass distributions]{\label{fig:emuMass}
    \includegraphics[width=0.48\textwidth]{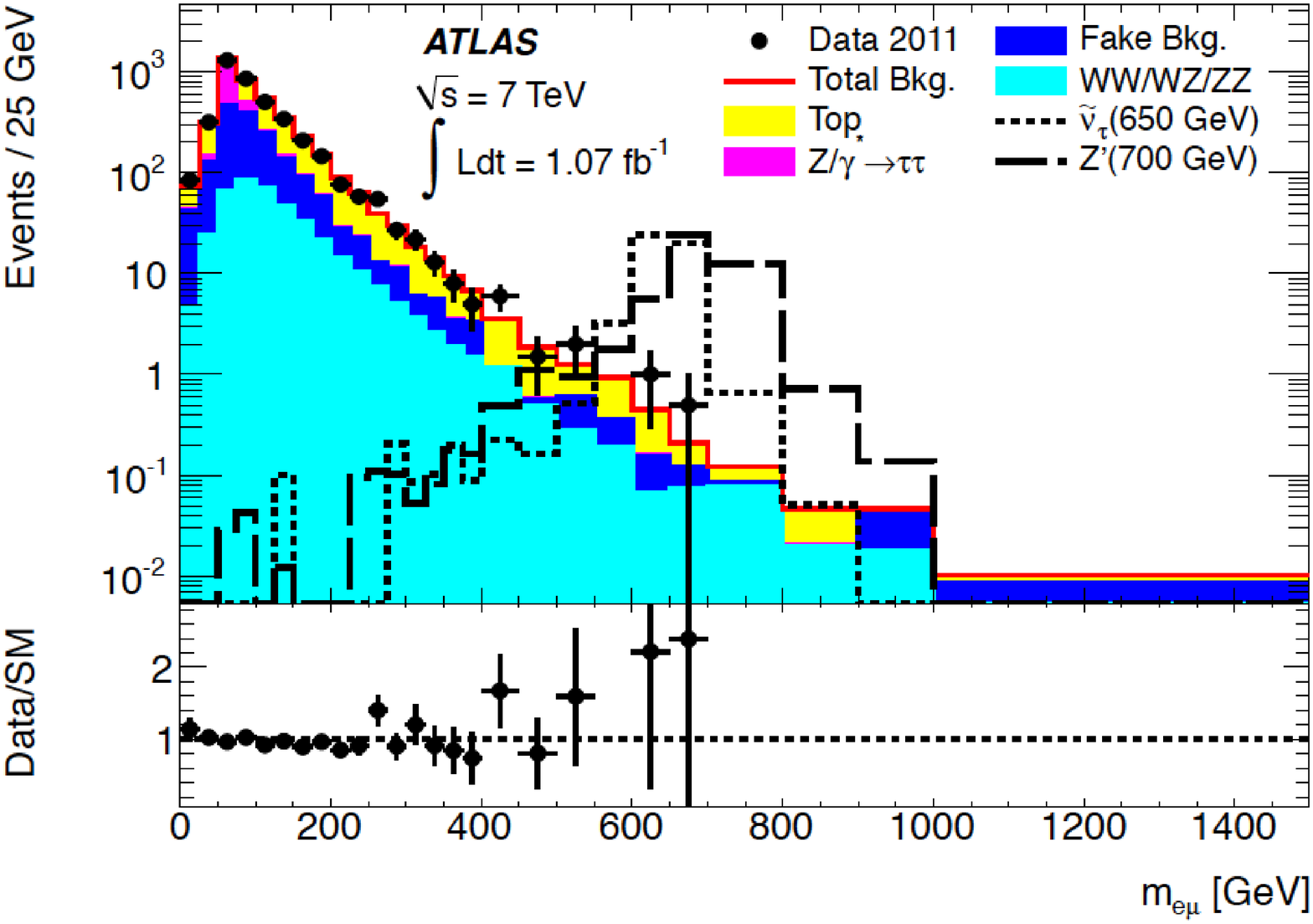}}
  \subfigure[95$\%$ C.L. upper limits on the $\lambda^{'}_{311}$ coupling as a function of $m_{\tilde{\nu}_{\tau}}$.]{\label{fig:emuLimits}
    \includegraphics[width=0.48\textwidth]{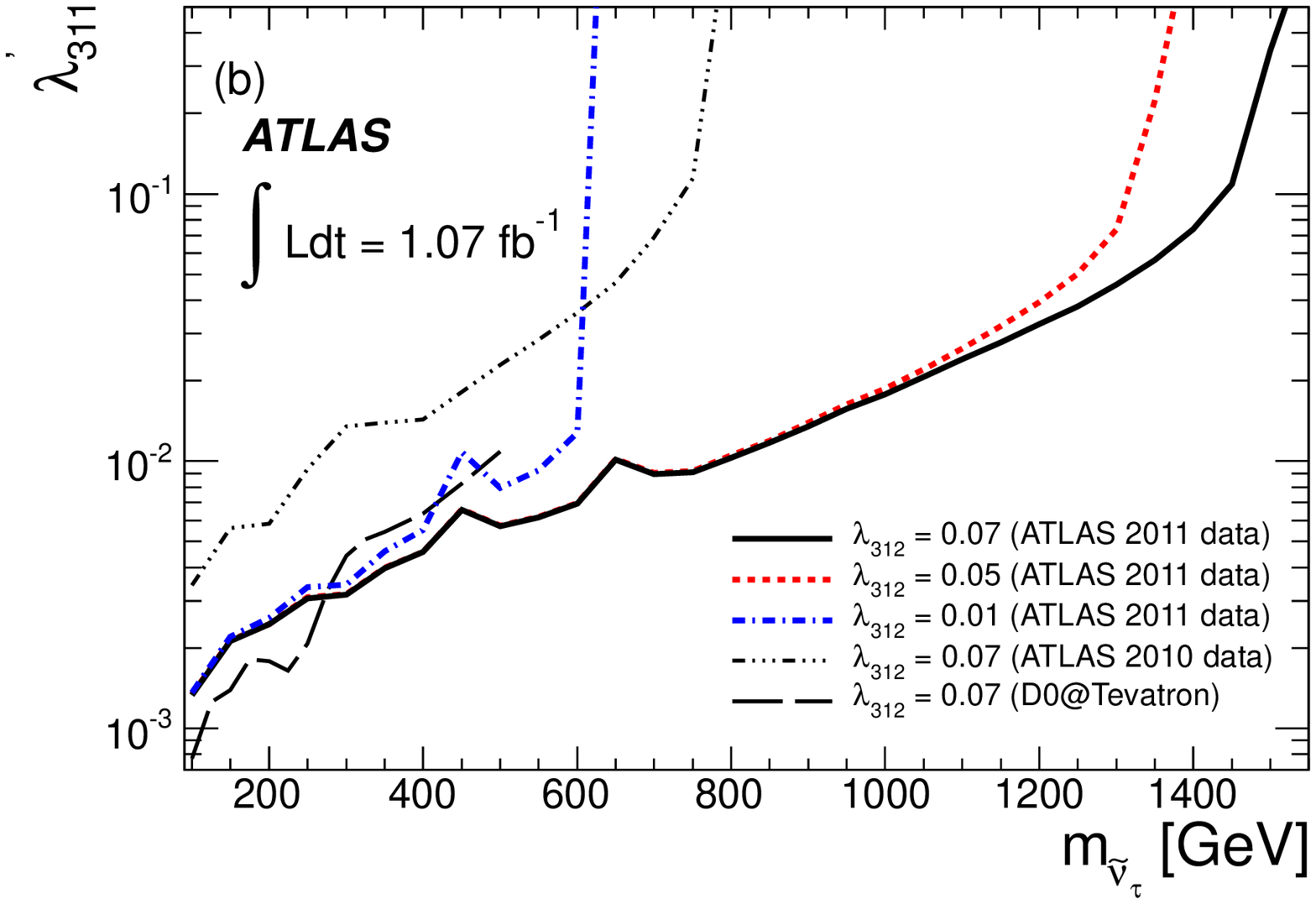}}
 \caption{Distribution of events and constraints on the RPV couplings for the $e\mu$ resonance search.}
\label{fig:emuSearch}
\end{figure}

Analysing 1.07$\pm$0.04~fb$^{-1}$ of data collected in 2011 using single lepton triggers (measured to be 100$\%$ efficient), 
a total of 4053 $e\mu$ candidates are observed, while 4150$\pm$250 are expected from Standard Model processes.
The distribution of the invariant mass $m_{e\mu}$ is presented in Figure~\ref{fig:emuMass}.  

Since no evidence for signal is observed the number of events in each search region, as a function of increasing
$m_{e\mu}$, are used to set an upper limit on 
$\sigma(pp\rightarrow\tilde{\nu}_{\tau})~\times$~BR$(\tilde{\nu}_{\tau}\rightarrow e\mu)$ using a Bayesian method
with a uniform prior for the signal cross section. Figure~\ref{fig:emuLimits} 
shows the 95$\%$ C.L upper limits on the $\lambda^{'}_{311}$ coupling as a function of $m_{\tilde{\nu}_{\tau}}$
for three values of $\lambda_{312}$. The regions above the curves are excluded in each case.  
Please refer to~\cite{bib:emuPaper} for the full limits and exclusion curves on the $\tilde{\nu}_{\tau}$ production
cross section.

%

\section{Displaced vertices}

We report on results of a search for a heavy particle decaying into multiple charged particles 
at a distance of order millimeters to tens of centimeters from the $pp$ interaction point
in events containing a muon identified with high $p_{\mathrm{T}}$~\cite{bib:dvPaper}.
In the SUGRA scenario, such a signature may be manifested via the decay of the lightest
supersymmetric particle due to non-zero $\lambda^{'}_{2ij}$ couplings via a diagram such as is shown in 
Figure~\ref{fig:dispVert}. Current limits on RPV couplings~\cite{bib:sugra} allow for the decay vertex 
to be displaced and within range of the ATLAS inner tracking detectors.

Events are selected from a data sample of 36~pb$^{-1}$ collected in 2010 and must
pass the $p_{\mathrm{T}} >$~40~GeV single-muon trigger requirement. 
A primary vertex (PV), originating from the $pp$ collision is required and must contain a minimum
of five tracks and a $z$ position within 200~mm. 
Where multiple primary vertices are recorded the one
with the highest scalar sum of the $p_{\mathrm{T}}$ of its tracks is used.

We reconstruct a displaced vertex by selecting only tracks with $p_{\mathrm{T}} >$~1~GeV. A large impact parameter ($>$~2~mm),
with respect to the transverse position of the PV is required, rejecting 98$\%$ of all tracks orginiating from the 
primary $pp$ interaction. A description of the displaced vertex finding algorithm can be found elsewhere~\cite{bib:dvPaper}.

\begin{figure}[hbtp]
\begin{center}
  \includegraphics[width=9cm]{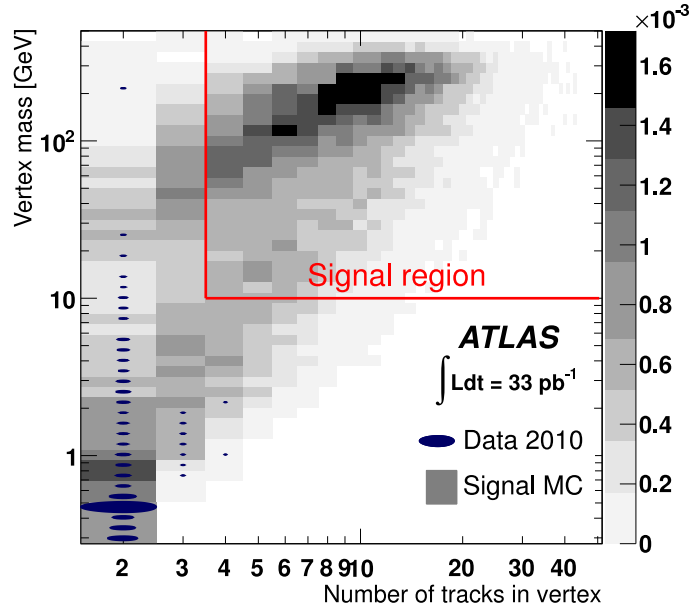}
  \caption{\label{fig:massVsMult} Vertex mass (\mdv) vs. vertex track
  multiplicity (\ntdv) for displaced vertices that pass 
  the event selection requirements except the
  \mdv\ and \ntdv\ requirements.
  Shaded bins show the distribution for signal MC, and
  data are shown as filled ellipses, with the area of the ellipse
  proportional to the number of events in the corresponding bin. 
  The figure contains 487 data events, of which 251 are in
  the bin corresponding to $K^{0}_{s}$ decays.}
\end{center}
\end{figure}

We use the mass of the reconstructed vertex ($\mdv$) and the number of reconstructed charged tracks ($\ntdv$) as discriminating 
variables for the search.
Figure~\ref{fig:massVsMult} shows the distribution of 
\mdv\ vs. \ntdv\ for the selected vertices in the data sample,
including vertices that fail the requirements on \mdv\ and \ntdv\,
this is overlaid with a distribution from signal MC. 
We observe no vertices that satisfy all the selection criteria.

%
%
%
\begin{figure}[hbtp]
\begin{center}
  \includegraphics[width=9cm]{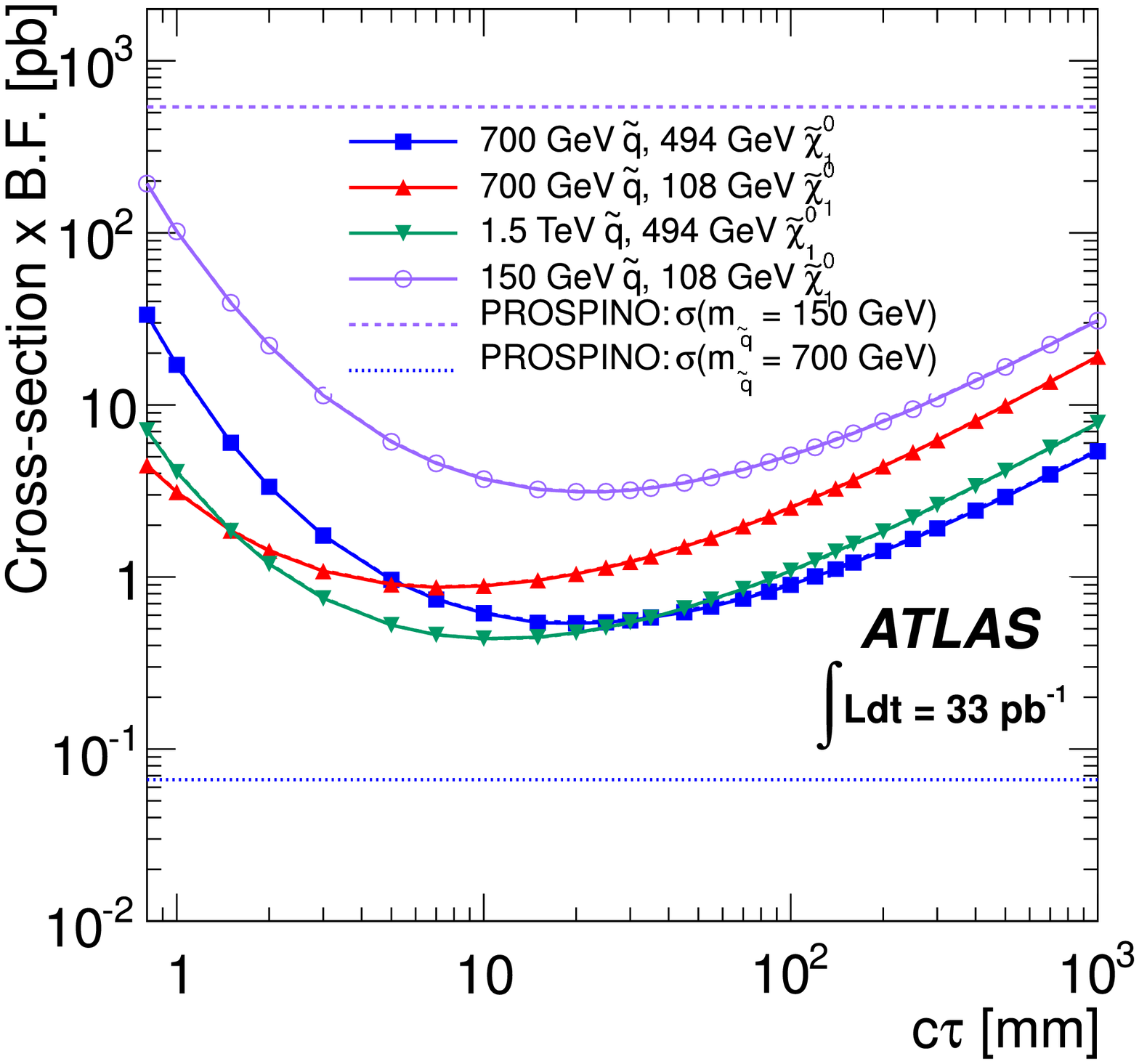}
  \caption{\label{fig:limits} Upper limits at 95\% CL on the production
  cross-section times branching fraction vs. the neutralino lifetime
  times the speed of light for different combinations of squark and
  neutralino masses, based on the observation of zero events in a
  \lumi\ data sample. The horizontal lines show the cross-sections
  calculated from PROSPINO~\cite{bib:prospino} for squark masses of 700~GeV and 150~GeV.
}
\end{center}
\end{figure}

Based on this null observation, we set upper limits on the
supersymmetry production cross-section times the branching
fraction of the simulated signal decays for different
combinations of squark and neutralino masses, and for different values
of the product of the speed of light ($c$) and the neutralino lifetime ($\tau$).
These constraints can be seen in Figure~\ref{fig:limits}.

\section{Summary}

The ATLAS collaboration has begun mining the data collected during the first two years of 7~TeV operations
at the CERN Large Hadron Collider. Novel analyses have already reached maturity, and searches for SUSY
with R-parity violation with two different approaches have been presented herein. At time of writing no
deviations from the Standard Model have been observed in these channels. However, these early analyses will form the 
foundation of further searches to be performed in the near future.

\end{document}